\documentclass[aps,amsmath,showpacs,amsfonts,twocolumn,10pt]{revtex4}
\usepackage{epsfig,graphicx}
\usepackage{setspace} 
\usepackage[english]{babel}
\usepackage{amsfonts}
\usepackage{amsmath}
\usepackage{latexsym}
\usepackage{graphics,bm}
\usepackage{natbib}
\usepackage{dcolumn}
\usepackage{bm}
\usepackage{rotating}
\usepackage{amsmath}
\usepackage{braket}
\usepackage{epstopdf}

\begin{document}

\title{Non equilibrium dynamics of an Optomechanical Dicke Model}

\author{Kamanasish Debnath$^1$}
\author{Aranya B Bhattacherjee$^2$}
\address{$^1$Institute of Applied Sciences, Amity University, Noida - 201303 (U.P.), India \\
 $^2$School of Physical Sciences, Jawaharlal Nehru University, New Delhi- 110067, India}

\begin{abstract}
Motivated by the experimental realization of Dicke model in optical cavities, we model an optomechanical system consisting of two level BEC atoms with transverse pumping. We investigate the transition from normal and inverted state to the superradiant phase through a detailed study of the phase portraits of the system. The rich phase portraits generated by analytical arguments display two types of superradiant phases, regions of coexistence and some portion determining the persistent oscillations. We study the time evolution of the system from any phase and discuss the role of mirror frequency in reaching their attractors. Further, we add an external mechanical pump to the mirror which is capable of changing the mirror frequency through radiation pressure and study the impact of the pump on the phase portraits and the dynamics of the system. We find the external mirror frequency changing the phase portraits and even shifting the critical transition point, thereby predicting a system with controllable phase transition.
\end{abstract}

\maketitle
\section{Introduction}

	Cavity optomechanics \citep{7} has been playing an important role in the exploration of quantum mechanical systems, especially the coupling between the electromagnetic field of the cavity and the mechanical oscillator \citep{1,2,3}.  The photons inside the ultrahigh finesse cavity are capable of displacing the mechanical mirror through radiation pressure and this has been a subject of early research in the context of nanomechanical cantilevers \citep{4,5,6}, vibrating microtoroids \citep{8}, membranes and Bose-Einstein condensates \citep{9}. Recent advancements in the field of laser cooling, high finesse nanomechanical mirrors have made it possible to study ultra cold atoms by combining the tools of cavity quantum electrodynamics. Experimental realisation of quantum entanglement, gravitational wave detection \citep{10,11} in the last few years has added new interest to the field of optomechanics. Such a system with an ensemble of N atoms with single optical mode has been an interesting theme in quantum optics after the work of Dicke \citep{12}, showing the effects of quantum phase transition and superradiant phases.The phase transition from a super fluid to a self organised phase, above a certain threshold frequency, when a laser driven BEC \cite{46, 47, 48, 49} is coupled to the vacuum field of the cavity refers to the basic Dicke model \citep{35, 36, 37}. The ultra cold atoms self organizes to form a checkerboard pattern trapped in the interference pattern of the pump and the cavity beams \citep{13,14,15}. This self organization initiates at the onset of the superradiance in an effective non equilibrium Dicke model. Since then many theoretical proposals for single mode, multi mode \citep{38} and optomechanical Dicke models has been made which are presumed to exhibit interesting physics \citep{39} and applications in the field of quantum simulation and quantum information \citep{40, 41, 42, 43}. In the present cold atom settings, the splitting of the two distinct momentum states of the BEC is controlled by the atomic recoil energy, and this enables the phase transition to be observed with optical frequencies with light. This is quite similar to the theoretical approach proposed by Dimer $\textit{et al.}$ \citep{17} for attaining Dicke phase transitions using Raman pumping schemes between the hyperfine levels \citep{44}.\\

	In this paper, we propose an optomechanical system consisting of N, two level elongated cigar shaped BEC interacting with light in a high finesse optical cavity with a movable mirror. Such systems can be used to investigate the optomechanical effects on the second order phase transition to a superradiant regime. We study the dynamics of the system and bring out all the possible phases  by analytical arguments and further propose a modification in the system that can be used to alter the phase portraits and transition point of the system.

\section{The Model}

We consider a Fabry- Perot optical cavity with one fixed and another movable high finesse mirror of mass $M$, capable of oscillating freely with frequency $\omega_m$. A two level, cigar shaped BEC is trapped within the cavity with transition frequency $\omega_a$. The optical cavity is subjected to a transverse pump beam with Rabi frequency $\Omega_p$, wave vector $k$ and frequency $\omega_p$. In order to avoid population inversion, the later is far detuned from the atomic transition $\omega_a$. Absorption and emission of cavity photons generates an effective two level spin system with spin down and spin up corresponding to the ground $\Ket{0,0}$ and excited states $\Ket{\pm k, \pm k}$ respectively. The effective Hamiltonian of such a system can be written as ($\hbar$= 1 throughout the paper) \citep{18, 21, 22}:- 

\begin{eqnarray}  
H&=& \omega_a S_z+ \omega a^{\dagger} a+ \omega_m b^{\dagger} b+ \delta_0 a^{\dagger} a (b+ b^{\dagger}) \nonumber\\
&+& g(a+ a^{\dagger})(S_{+}+ S_{-})+ US_z a^{\dagger}a,
\end{eqnarray}

\begin{figure}[h!]
\includegraphics[width=0.45\textwidth]{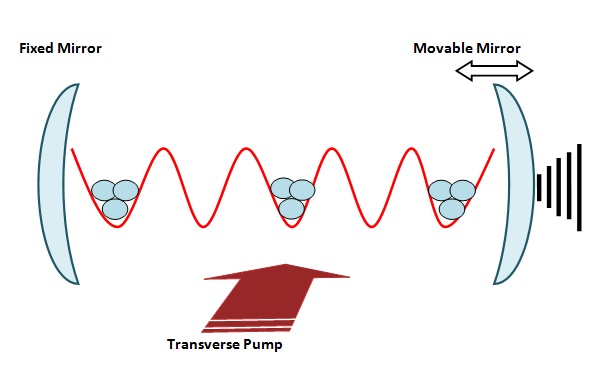}
\caption{The schematic representation of the model considered. One of the mirror is movable under the radiation pressure of the cavity beams. The optical cavity has a decay rate of $\kappa$ and the mechanical mirror has a damping rate $\Gamma_m$.}
\centering
\end{figure}	

where $\omega= \omega_c- \omega_p- N(5/8)g_0^2/ (\omega_a- \omega_c)$ \citep{22}, $\omega_c$ being the cavity frequency. $U$ represents the back reaction of the cavity light on the BEC and is given by $U= -(1/4)g_0^2/ (\omega_a- \omega_c)$, which is generally negative, however both the signs are achievable experimentally and we shall deal with the both in the present paper. $\delta_0$ is the optomechanical single photon coupling strength which represents the optical frequency shift produced by a zero point displacement. $\delta_0$ can be identified as $\omega x_{ZPF}/{L}$, $L$ being the cavity length and $x_{ZPF}$ denoting the mechanical zero point fluctuations (width of the mechanical ground state wave function) \citep{50}. $\omega_m$ represents the frequency of the mechanical mirror, which generates phonons with $b (b^{\dagger})$ as the annihilation (creation) operator. In the experiments of \citep{14}, both the pump and cavity were red detuned from the atomic transition and hence $U$ was considered negative for the observed Dicke phase transition. $a (a^{\dagger})$ is the annihilation (creation) operator of the optical mode while $b (b^{\dagger})$ representing the same for the mechanical mode, following the commutation relation [$a (b), a^{\dagger} (b^{\dagger})]$= 1. $S_{+}, S_{-}$ and $S_z$ are the spin operators obeying the relation [$S_+, S_-$]= $2S_z$ and [$S_{\pm}, S_z$]= $\mp S_{\pm}$. $\textbf{S}= (S_x, S_y, S_z)$ is the effective collective spin of length $N/2$. The co and counter rotating matter light coupling has been taken equal throughout the paper and is denoted by $g$. The schematic representation of the model considered in this paper has been shown in Fig. (1). \\

In the thermodynamic limit, the semi classical equations for our system, takes the form: -

\begin{equation}
\dot{S_{-}}= -i(\omega_a+ U \mid a \mid ^2) S_{-}+ 2ig(a+ a^{\dagger})S_z,
\end{equation}

\begin{equation}
\dot{S_z}= -ig(a+ a^{\dagger})S_{+}+ ig(a+ a^{\dagger})S_{-},
\end{equation}

\begin{eqnarray}
\dot{a}&=& -[\kappa+ i(\omega+ US_z+ \delta_0 (b+ b^{\dagger})]a\nonumber\\
&-& ig(S_{+}+ S_{-}),
\end{eqnarray}

\begin{equation}
\dot{b}= -i\omega_m b- i\delta_0 \mid a \mid^2- \Gamma_m b,
\end{equation}

where $\kappa$ and $\Gamma_m$ are the cavity decay rate and damping rate of the mechanical mode respectively. We employ the steady state analysis $(\dot{S}_-= \dot{S_z}= \dot{a}= \dot{b}= 0)$ of the above equations to determine the critical atom- cavity coupling strength. We carry a numerical approach in this paper to determine the critical value $\lambda_c$ ($g\sqrt{N_c}$). The analytical process uses the c- number variables and quantum fluctuations, and one can refer \citep{21} for the complete process in the absence of back reaction term. $\lambda$ ($g\sqrt{N}$) $>$ $\lambda_c$ ($g\sqrt{N_c}$) marks the onset of the superradiance, which was first observed experimentally by Tilman Esslinger and his group \citep{14} for BEC atoms in 2010.

\section{Superradiant phases and  Phase Portraits}

To study the dynamics of the present system, we employ the same mathematical technique as \citep{22, 27} and define $a= a_1+ ia_2$, $b= b_1+ ib_2$ and $S_{\pm}= S_x \pm iS_y$. Substituting the same in the above semi classical equations (Eq. (2)- (5)) and comparing the real and imaginary parts on both side yields: -

\begin{equation}
(\omega_a+ U\mid a \mid^2)S_y= 0,
\end{equation}

\begin{equation}
(\omega_a+ U\mid a \mid^2)S_x- 4ga_1S_z= 0,
\end{equation}

\begin{equation}
-\kappa a_1+ (\omega+ US_z+ 2\delta_0 b_1)a_2= 0,
\end{equation}

\begin{equation}
\kappa a_2+ (\omega+ US_z+ 2\delta_0 b_1)a_1+ 2gS_x= 0.
\end{equation}

\begin{equation}
b= -\Big(\frac{\delta_0 \mid a \mid ^2 \omega_m}{\Gamma_m^2+ \omega_m^2}\Big) - i\Big(\frac{\delta_0 \mid a \mid^2 \Gamma_m}{\Gamma_m^2+ \omega_m^2}\big)
\end{equation}

Clearly, from Eq. (6), either $S_y$= 0 or $(\omega_a+ U\mid a \mid^2)$= 0. We define the case arising from the first condition as the superradiant phase A (SRA) and the second condition as the superradiant phase B (SRB). SRA represents the quantum phase transition from normal (N) or inverted (I) states to a self organized states defined by $S_x$ and $S_z$ only. Similarly, the SRB represents the transition from the mixed states (N+ I) to a superradiant phase defined by all the components of $\textbf{S}$. The difference of transition from mixed states (N+ I) as in SRB phase compared to from normal (N) or inverted (I) as in SRA phase can be understood in the phase diagrams. Ofcourse, with increasing back action parameter, we expect a reduced phase transition region. Again, SRB phase condition limits $U$ to be only negative, since the phase is defined as $(\omega_a+ U\mid a \mid^2)$= 0.  However, what might be the effect of the mechanical mirror motion on the phase transition of the system? In the absence of the back reaction parameter $U$, \citep{21, 27} suggests no change in the critical transition point, $\lambda_c$ for the SRA phase. However, in the presence of the back reaction term and in the SRB phase, what role can the mirror frequency play in defining the phase portraits, a question to be analyzed in this paper. In the next section, we shall analyze all the possible conditions and present the phase portraits of the system for both positive and negative back reaction parameter.

\subsection{SRA Phase}

As defined before, $S_y$= 0 marks the SRA phase, which is simply the transition from normal (N) or inverted (I) state to the regime of superradiance. The critical atom- cavity coupling point can be determined by setting $[S_x, S_y, S_z]= [0, 0, \pm N/2]$, which signifies the presence of either spin up (inverted) or spin down (normal) particles and no photons. The steady state equations (Eq. 6- 9) can be straightforwardly solved using matrix method for $S_z$, which yields a quadratic equation supporting two roots of $S_z$. The determinant representing the steady state equations, takes the form: -

\begin{equation}
\begin{vmatrix} 
\omega_{a} + U\mid a \mid^2 & 0 & -4gS_{z} & 0 \\ 0 &  \omega_{a} + U\mid a \mid^2 & 0 & 0 \\ 2g & 0 & \chi & \kappa  \\ 0 & 0 & -\kappa & \chi \end{vmatrix}= 0, 
\end{equation}

where $\chi=\omega+ US_z- \frac{2\delta_0^2\mid a \mid^2 \omega_m}{\Gamma_m^2+ \omega_m^2}$. The above determinant has been solved numerically and the results are too cumbersome to be reproduced here. The two supporting roots for $S_z$ when equated to $\pm N/2$, and solved for $\omega$, represents the dynamical phase portrait for SRA phase showing the transition from normal (N) and inverted (I) phase to regimes of superradiance. An important point to note here, is that the SRA phase exists for any value of the back action parameter, U. Although the two roots of $S_z$ must be independent, however, we shall find a small region in the phase portraits, where both the roots of $S_z$ are satisfied. Such regions has been defined as 2SRA phase, or more precisely as SRA (N)+ SRA (I) phase. The same also had its existence in \citep{22}, however, in this paper, we shall find the mirror frequency $\omega_m$ to determine the physics of such coexisting regime and we shall exploit such condition to alter the phase portraits.

\subsection{SRB phase}

We defined the condition $(\omega_a+ U\mid a \mid^2)$= 0 as the origin of the B type superradiance. The same condition when incorporated in Eq. (7), yields $4gS_z a_1$= 0. Evidently, this bounds $a$ to be purely imaginary. Correspondingly, the initial condition also yields:-

\begin{equation}
\mid a \mid ^2= -\frac{\omega_a}{U},
\end{equation}

which again suggests the same nature for $a$. Hence $a_1$= 0, which when plugged in Eq. (8) and Eq. (9) yields: -

\begin{equation}
S_x^2= -\frac{\kappa^2 \omega_a}{2gU}
\end{equation}

and

\begin{equation}
S_z^2= \Big(\frac{\omega+ 2\delta_0 b_1}{U}\Big)^2.
\end{equation}

As noted previously, $S_y$= 0 was defined in SRA phase and in SRB phase $S_y$ $\ne$ 0. Hence, it follows from the normalization condition that $S_x^2+ S_z^2 \le$ $N^2/4$, where the above expressions give the corresponding values, with Eq. (10) determining the expression for $b_1$ and $\mid b \mid^2$.

\begin{figure}[h!]
\includegraphics[width=0.45\textwidth]{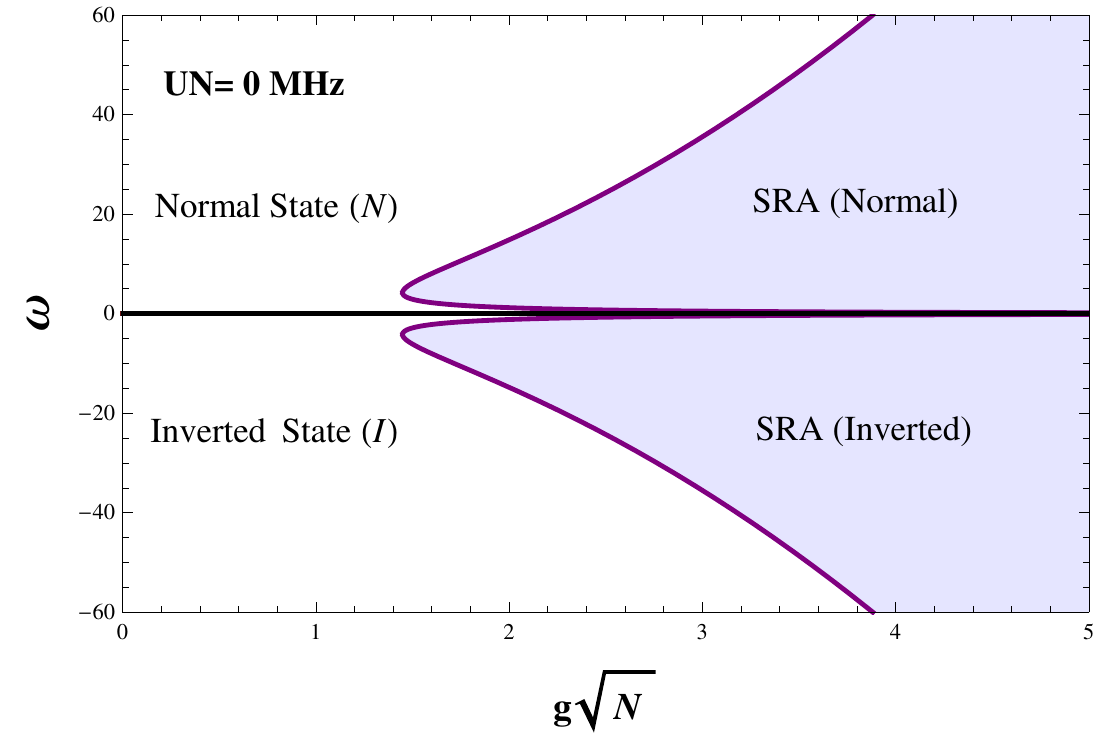}
\includegraphics[width=0.45\textwidth]{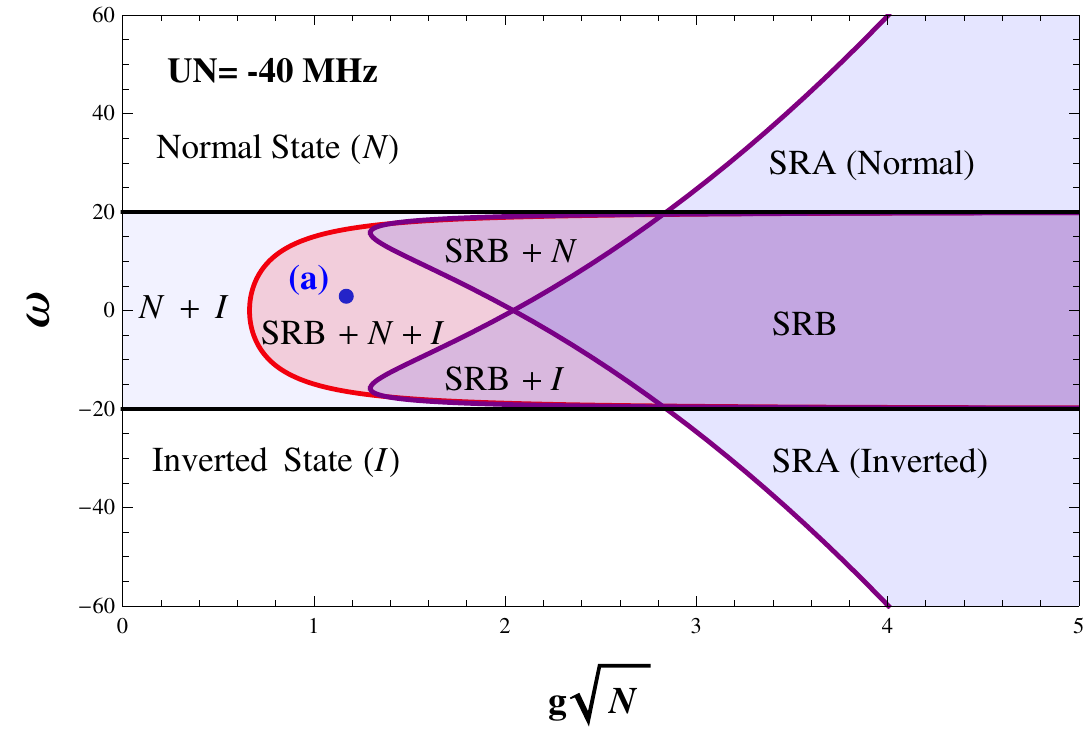}
\includegraphics[width=0.45\textwidth]{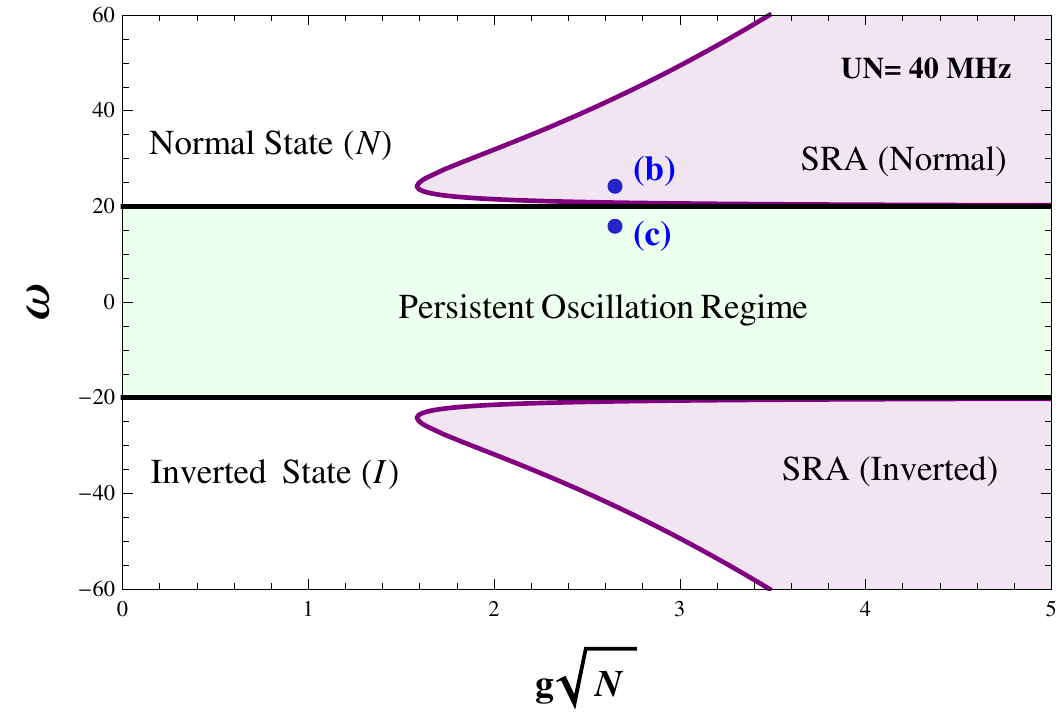}
\caption{Dynamical phase portrait of the stable attractors as a function of cavity frequency $\omega_c$ and $g\sqrt{N}$. The panels represent plots for $UN$= 0 MHz, -40 MHz and 40 MHz. Other parameters chosen were $\omega_m= 1$, $\omega_a= 0.05$ MHz \citep{14}, $\kappa$= 8.1 MHz \citep{14, 22, 23} and $\delta_0$= 0.05 and $\Gamma_m$= 0.05 $\omega_m$ \citep{21}.}
\centering
\end{figure}

\subsection{The Phase portraits}

We finally summarize the phase portraits of the dynamical system, with chosen parameters that satisfy the Routh- Hurwitz criteria \citep{33, 34} for a stable optomechanical system. We plot the phase portraits as a function of $g\sqrt{N}$, where N is the number of atoms $\approx$ $10^6$. We consider all the cases possible through analytical treatment of the dynamical equations of the system and it is noteworthy to mention here that although all these phase regions can be investigated in various experimental conditions, however, not all will emerge in a single experiment. The designing of such a system to observe various phase regions discussed here is a matter of technological advancement in controlling the parameters of the system. Experiments reported by K. Baumann $\textit{et al.}$ \citep{14, 15} showed the system evolving from normal phase (N) with all spins pointing downwards and no photons.

The first panel of fig.2 shows the phase diagram for UN= 0 MHz. The purple line marks the onset of superradiance from both normal (N) and inverted (I) states with all spins pointing downwards and upwards respectively. For $\omega<$ 0, the normal state (N) becomes unstable and the inverted state (I) becomes stable instead. As the backaction parameter is reduced (UN= -40 MHz), the SRA phase boundaries (purple line) between the (N) and (I) state shift to higher and lower frequency respectively. Simultaneously, SRB phase (red line) emerges which coincides with SRA(N) and SRA(I) for negative U as discussed previously and few new regimes come to play as seen from the second panel of fig. 2. The (N) and (I) phase coexists due to the shift of the SRA boundary and also gives rise to (SRB+ N), (SRB+ I) and (SRB+ N+ I) regions. Due to the frequency shifts induced by negative U, there exists a small region where SRA(N) and SRA(I) coexists, where both the roots of $S_z$ are supported. These phases are represented as 2SRA (SRA(N)+ SRA(I)) in this paper and we shall deal with the same in next section. \\

	Although we have portrayed all the possible cases (for $UN$= -40 MHz) in the middle panel of Fig. (2), not all can be simultaneously observed in any single experiment. As reported by Esslinger and his group \citep{14}, the first superradiant transition was observed from inverted state (I) to SRA (Inverted) which corresponds to the lower symmetrical half of the phase portrait. The realization of other transitions is purely dependent on the conditions of the system. Considering $S_y$= 0 and the initial state being the normal state (-N/2 and no photons), the phase transition would correspond to the SRA (N) denoted by the purple line on the positive Y axis of top panel of Fig. (2) and vice versa for system prepared with inverted state (N/2) and operated with negative effective cavity frequency ($\omega$). The purple line marks the phase transition from superfluid to a self organized state and as seen from the figure, the critical transition point increases as the effective cavity frequency ($\omega$) in increased. This also supports the analytical results in \citep{14, 17, 21, 22, 23, 29} which showed the critical point at $\frac{1}{2} \Big( \frac{\omega_a}{\omega} (\kappa^2+ \omega^2) \Big)^{1/2}$ for $U$= 0. \\

	Interestingly, as we reduce the back reaction parameter $U$ (middle panel, Fig. (2)), the SRA phase boundary shifts towards each other by $\pm UN/2$ so as to offer an identical superradiant phase (area covered between the purple lines and black horizontal lines (at $\pm UN/2$)  in the middle panel of Fig. 2). Although we can never witness both the transitions in a single experiment, however, theoretical study predicts an identical superradiant phase when operated with effective cavity frequency ranging between $\pm UN/2$ MHz and initial state being normal or inverted. In simple words, for $UN$= -40 MHz we can predict an identical phase transition when operated with $-20 \le \omega \le 20$ MHz without worrying whether we started from normal (N) or inverted (I) state. Thus we can start from any mixed state configuration ($i.e.$ a combination of spin up (I) and spin down (N)) and still expect to get a phase transition if $UN$ is negative. This is analogous to the case of preparing mixed atoms with 50$\%$ spin up and 50$\%$ spin down and still get an identical superradiance as in two atom Dicke model \citep{52}. The advantage lies in the fact that with negative back action parameter, we get a short window of selecting our effective cavity frequency ($- UN/2 \le \omega \le +UN/2$) and worry not about the initial condition (N or I) to observe a Type A superradiance. The comparison becomes evident when we see the top panel of Fig. (2), which showed phase transition only when the system is operated and prepared in a combination of either positive $\omega$ and Normal state (+$\omega$, N) or negative $\omega$ and Inverted state (-$\omega$, I). Thus a negative variation of $U$ gives us a freedom to choose our effective cavity frequency ($\omega$) and initial state. \\
	
	As the backaction parameter is made positive (UN= 40 MHz), the SRB phase vanishes for obvious reasons discussed previously. The SRA (N) and SRA (I) shifts away from each other by $\pm UN/2$ as seen from the lower panel of Fig. (2). The separation of the boundaries in opposite direction leads to the formation of another region termed here as persistent oscillation regime. Evidently, no phase transition can be observed when the system is operated with effective cavity frequency ($\omega$) between $\pm UN/2$. As the name suggests, this regime describes persistent oscillation and no steady state is reached even for long duration experiments, thereby predicting the presence of limit cycle. The notion of persistent oscillation will become clear in time evolution section when we shall simulate the system with initial conditions described by point (c), which lies in the concerned region.\\

	We observe the type B superradiance only when $(\omega_a + U\mid a \mid ^2)$=0 $i.e.$ when $U$ is negative since $\omega_a >$0. The critical line separating the superfluid and self organized state has been denoted with red colour in the middle panel of Fig. (2). The SRB imposing condition ($S_x^2+ S_z^2 \le N^2/4$) itself reveals the fact that it can take both $\pm N/2$ ($i.e.$ both Normal (N) and Inverted (I) state), which marks its appearance between $\pm UN/2$ in the phase portraits. Thus when $UN$= -40 MHz, and we have an initial mixed state configuration (N+ I) and effective cavity frequency $(\omega$) being operated between $\pm 20$ MHz, we can get either a Type A superradiance or a Type B superradiance depending on whether $S_y$= 0 or $(\omega_a+ U\mid a \mid^2)$= 0 respectively but never both simultaneously in a single experiment.

\section{2SRA phase}

\begin{figure}[h!]
\includegraphics[width=0.45\textwidth]{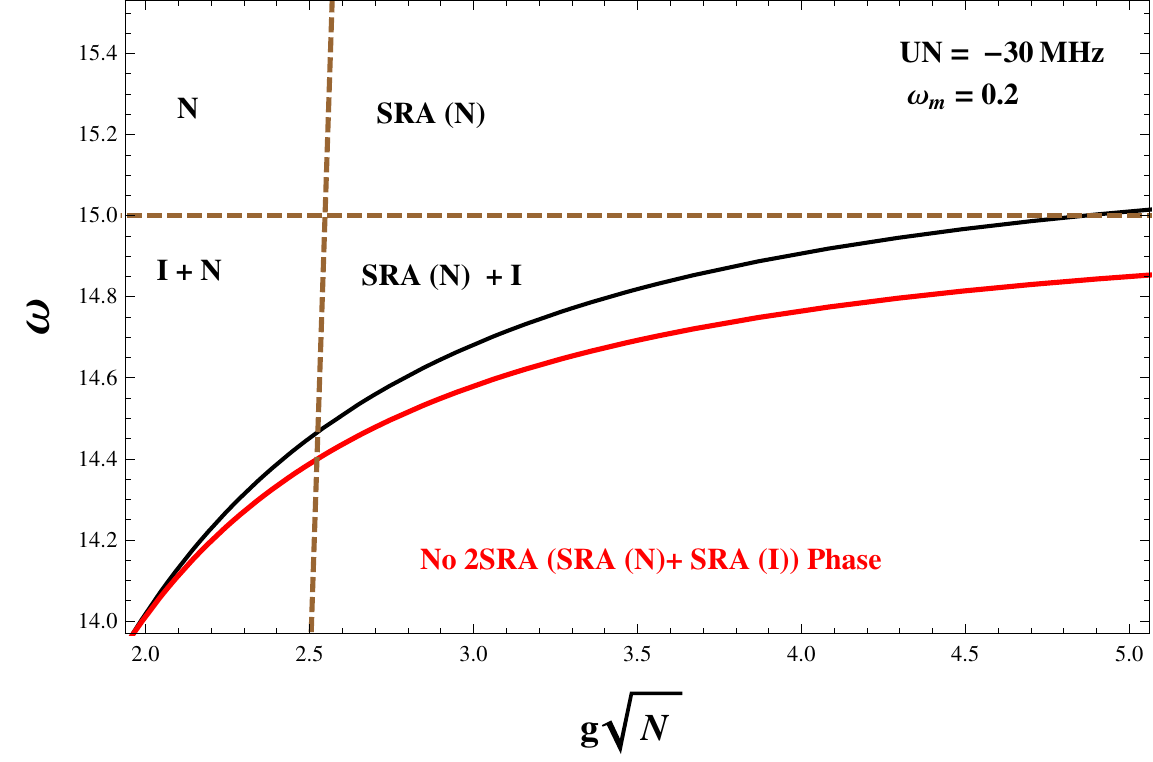}
\includegraphics[width=0.45\textwidth]{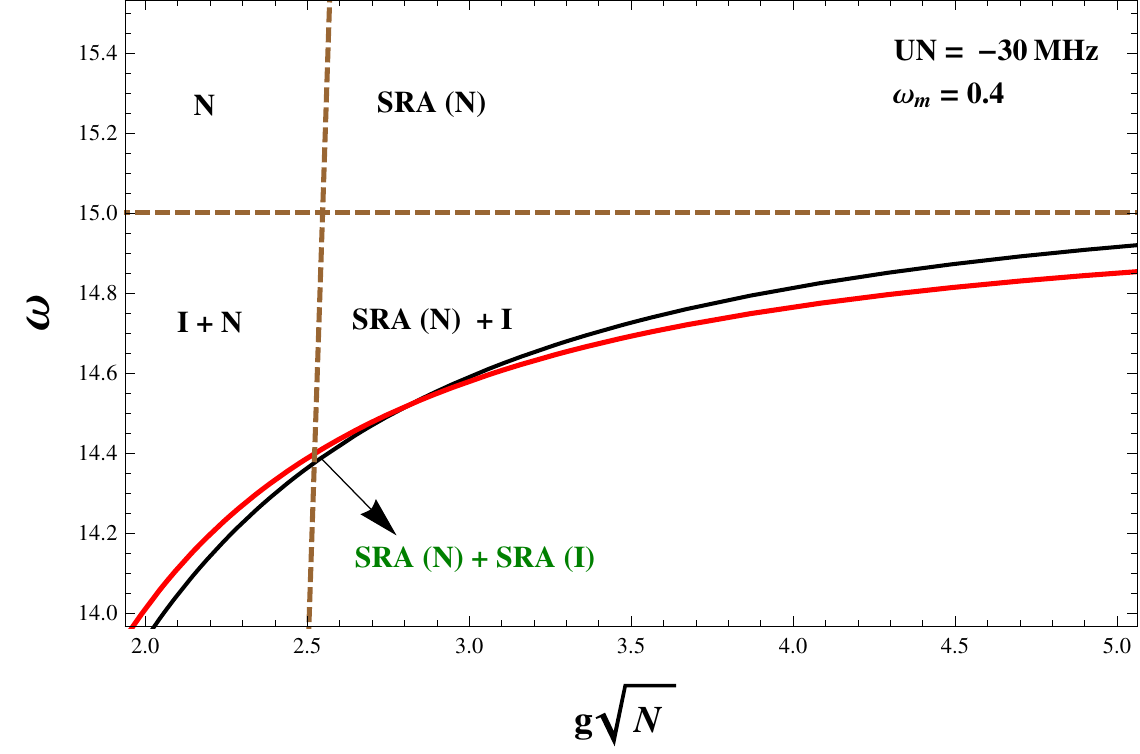}
\includegraphics[width=0.45\textwidth]{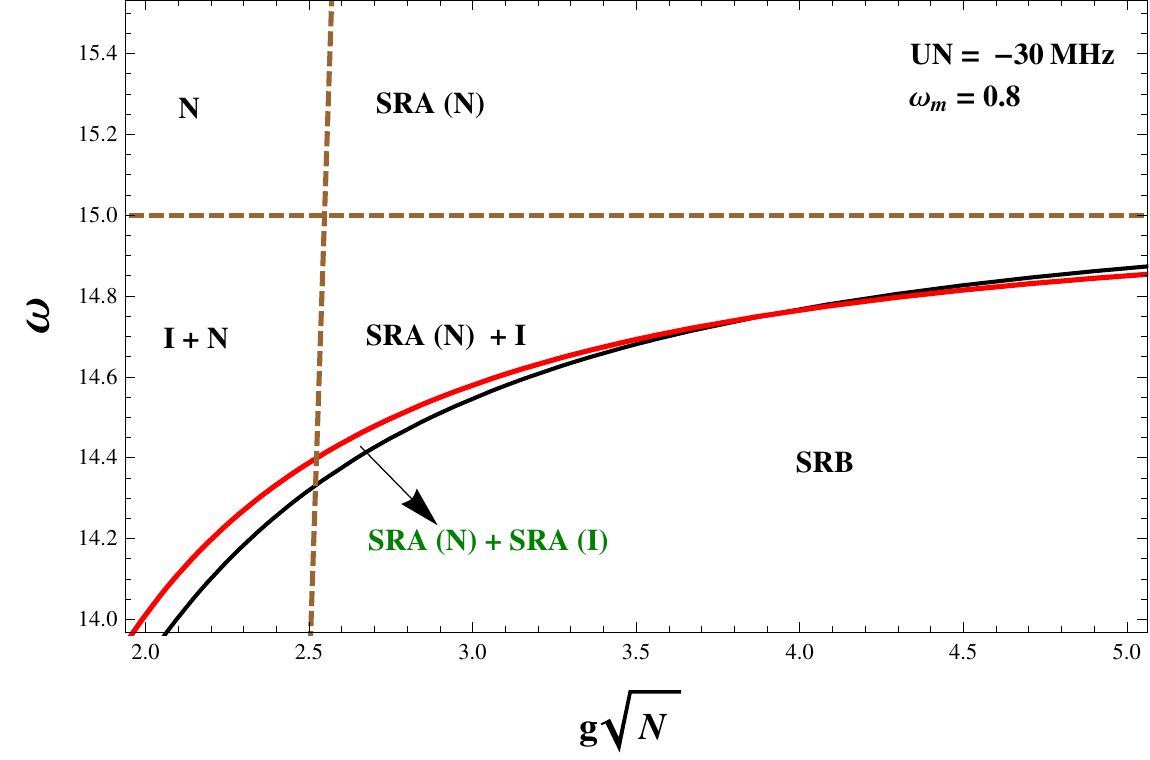}
\caption{Magnified view of the dynamical phase diagram for UN= -30 MHz for $\omega_m$= 0.2, 0.4 and 0.8 for upper, middle and lower panel respectively. Parameters chosen were same as in the previous plots.}
\centering
\end{figure}

In this section we aim to discuss the role of the mechanical mirror in defining the phase portraits of the system. As hinted previously, there are regions where both the roots of $S_z$ are supported and the SRA(N) and SRA(I) regions coincides to describe the new phase. Although evident from previous discussion that the mirror frequency plays no role in defining the SRA region, however, the SRB phase does have an explicit dependency on $\omega_m$ as seen from Eq. (13) and (14), together with Eq. (10) for the expression of $b_1$. We produce here a magnified view of the dynamical phase diagram for UN= -30 MHz and determine the variation in transition point for different values of $\omega_m$. Interestingly, the 2SRA phase is no more distinct as in the case of a fixed mirror \citep{22} and in the optomechanical case, the mirror frequency determines the physics of this tricritical point where all the phase boundaries cross each other. \\

For $\omega_m$= 0.2, the top panel of Fig. (3) shows no 2SRA region and the same starts becoming prominent as the mirror frequency $\omega_m$ is increased as seen from middle and the lowermost plots of Fig. (3). The mirror is therefore found to be altering the coexisting regime, for experimentally realizable values of the mirror frequency. These optical systems with a movable cantilever can therefore be efficiently used for controlling the crtitical point and also the coexisting regime. With these plots, the effect of the mirror frequency can be well established. However, we may demand to alter the phase portraits more since the change with the mechanical mirror is almost negligible for any use as in experimental phenomenon like quantum entanglement or manipulation etc. So can we devise and conceive any further modification to the system that can allow further manipulation of the critical transition point. We shall deal with the same in Sec VI, with an aim of modifying the phase diagrams by some easy controllable parameter.

\section{Time Evolution}

In order to get insight on the distinction between the described phases, we examine the time evolution of the system from various initial conditions lying in different phase regions. We mainly consider the points (a), (b) and (c) marked in the dynamical phase diagrams (fig 2) which lies in the (SRB+ N+ I), SRA and persistent oscillation regime respectively. We solve the semiclassical equations of the system numerically for $S_x$, $S_y$, and $S_z$  by fourth order Runge Kutta method and illustrate the relaxation time in reaching their corresponding stable attractors. The plots below shows the time evolution of the system from different initial conditions.

\begin{figure}[h!]
\includegraphics[width=0.45\textwidth]{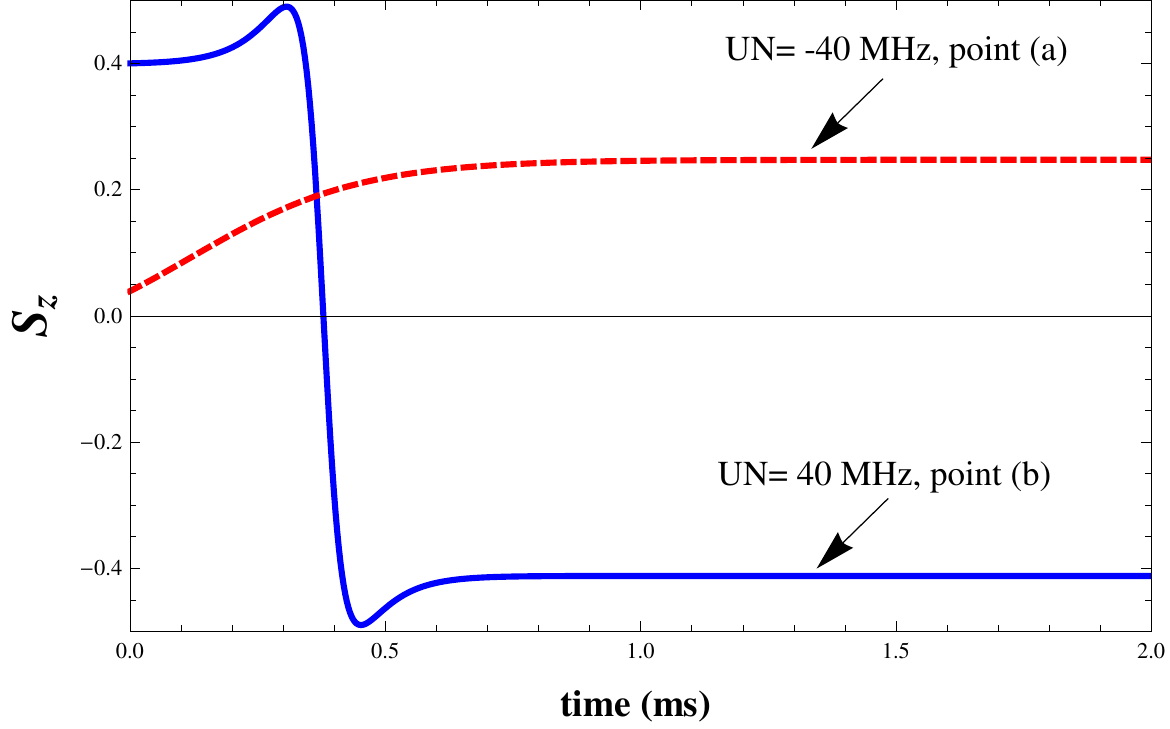}
\includegraphics[width=0.45\textwidth]{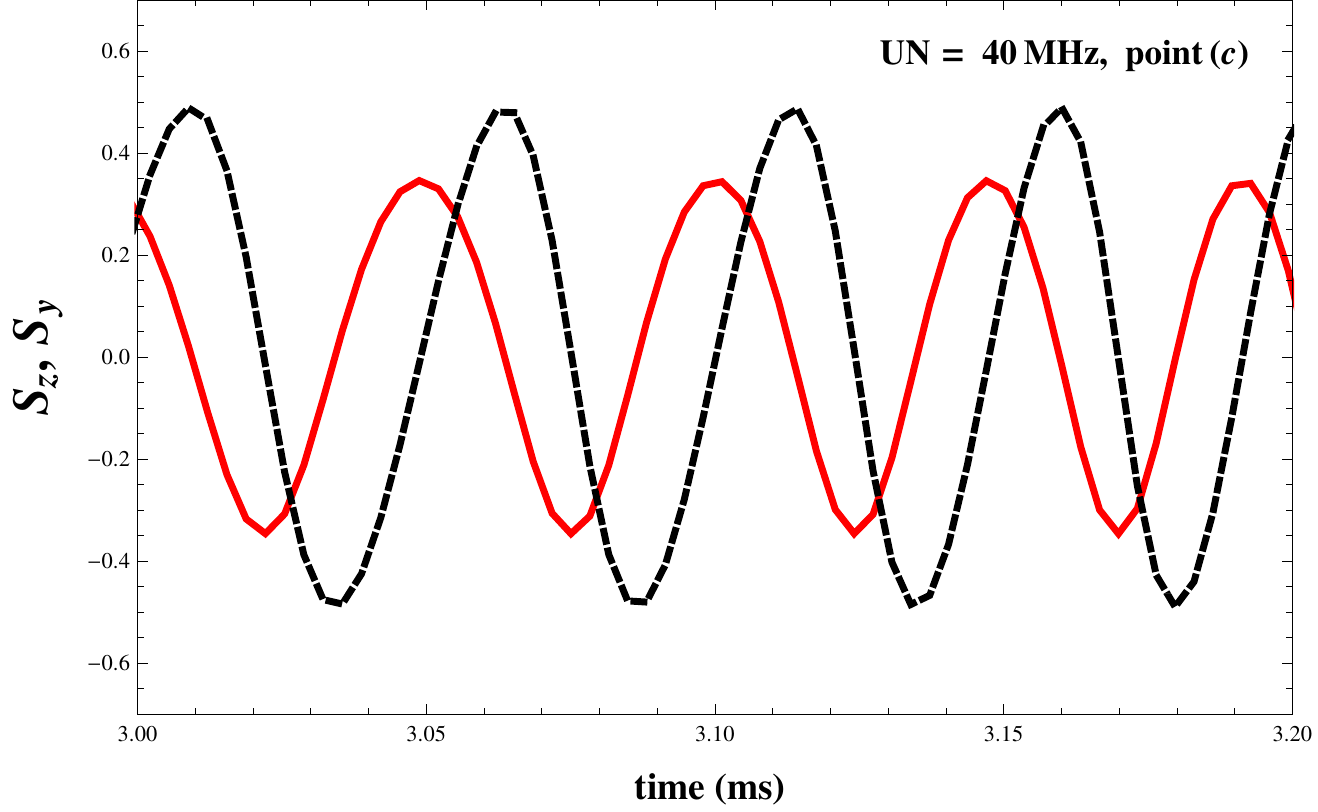}
\caption{Time evolution of the system from different initial condition. The top panel describes the point (a) and point (b) of fig.1 and the lower panel shows the persistent oscillations predicting a limit cycle in persistent oscillation regime for positive back action parameters. Other parameters used are same as in previous plots.}
\centering
\end{figure}

The top panel of fig. 4 shows the time evolution for point (a) (UN= -40 MHz) and point (b) (UN= 40 MHz) in the superradiant regime that are close to the normal and inverted state. The plot well describes the relaxation time for reaching their stable attractors. For point (a), $S_z$ initially increases and finally attains a stable value in approximately 0.7 ms thereby prediciting a stable case for realistic experiments. Point (b) lies in the SRA (N) region just above the persistent oscillation region and the time evolution of $S_z$ (blue curve) shows the system reaching their stable attractors in approximately 0.7 ms. As the initial condition enters the oscillation regime (point(c)), all the system parameters ($S_y$, $S_z$) starts to oscillate periodically at long times and no stable points are reached even after long duration as shown in the lower panel of fig. 4. Since the motion is described in a two dimensional plane, the attractors represent a simple limit cycle \citep{24}, thereby tagging the entire bounded plane as persistent oscillation regime.

\section{Dicke model in the presence of an external pump}

We modify our previous model by adding an external mechanical pump, which can be any external object in physical contact with the mirror or an external laser that is capable of changing the mirror frequency via radiation pressure. The pump can excite the mirror by coupling with the mirror fluctuation quadratures. The Hamiltonian of the new system, takes the form ($\hbar$= 1 throughout the paper) \citep{19, 27}: -

\begin{eqnarray} \label{eqn1}
H&=& \omega_a S_z+ \omega a^{\dagger}a+ \omega_m b^{\dagger}b+ \delta_0 a^{\dagger}a(b+ b^{\dagger}) \nonumber \\
&+& g(a+ a^{\dagger})(S_{+}+ S_{-})+ US_za^{\dagger}a+ \eta_p(b+ b^{\dagger}),
\end{eqnarray}

\begin{figure}[h!]
\includegraphics[width=0.45\textwidth]{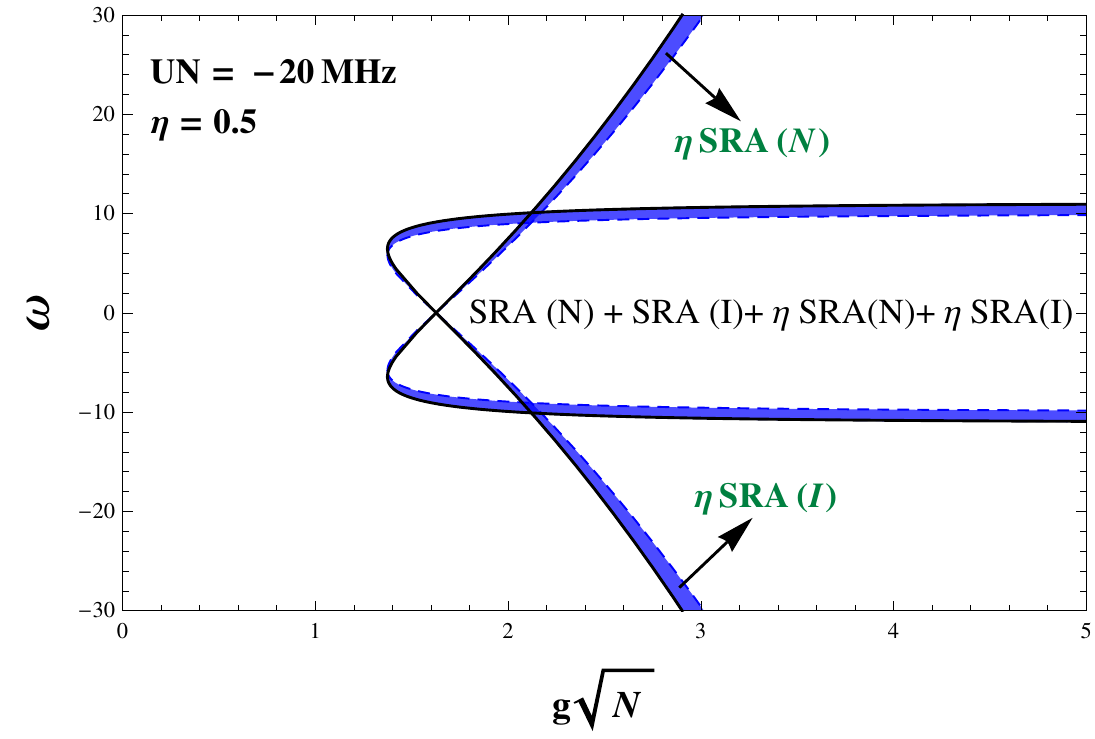}
\includegraphics[width=0.45\textwidth]{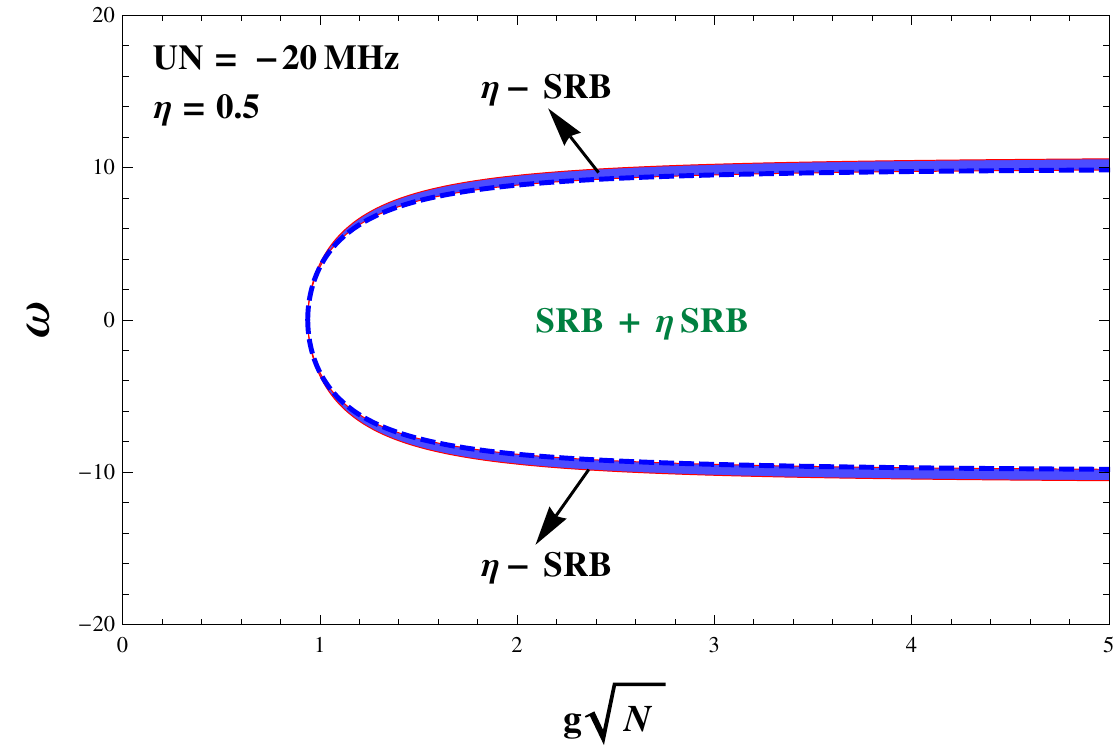}
\caption{The dynamical phase diagram in the presence of the external mechanical pump. UN= -20 MHz and $\eta_p$= 1. Other parameters are same as in previous plots.}
\centering
\end{figure}

where $\eta_p$ represents the mechanical pump frequency and the last additional term describes the energy due to it. The mechanical pump frequency will be considered to be small here, $\textit{i.e.}$ $ 0 \le \eta_p \le1$.  To proceed further, we begin with the semiclassical equations, of which the following equations gets modified: -

\begin{equation}
\dot{b}= -i\omega_m b- i\delta_0\mid a \mid^2- i\eta_p- \Gamma_m b.
\end{equation}

 We repeat the same analysis to determine the dynamical phase diagram of the system in the presence of the mechanical pump for both SRA and SRB phase. We produce here the dynamical phase portraits for SRA and SRB separately in fig. 5 to unveil the effect of the mechanical pump. The dotted lines in both the plots of fig. 5 marks the SRA and SRB phase boundaries in the absence of the mechanical pump and the bold curve represents the phase portrait when the external mechanical pump starts working. The blue shaded region named $\eta$- SRA and $\eta$- SRB represents the extra region created by the external mechanical pump. Clearly, the shaded region decreases the critical transition point both in positive and negative direction symmetrically. Although the external mechanical pump has changed the dynamical phase diagram to a large extend, the physics behind the time evolution remains almost same as in previous case with minor change in the relaxation time. The external mechanical pump frequency $\eta_p$  also enhances the 2SRA region to a large extend since both the SRA and SRB phase shows considerable increase in phase area. We don't produce these plots as these remains evident from the plots of fig. 4. Thus it is clear from the discussion in this section and section IV that the phase portraits can be altered and enhanced by a simple modification. Although the SRA phase region is unaltered by mirror frequency initially, the same can be modified when we add external force to the mirror. These systems can be used for altering the phase transition in Dicke model by simple controllable parameters like the external mechanical pump. Such can find use in experiments like detecting quantum entanglement, which tends to infinity at the critical point \citep{51}.

\section{Conclusion}

In this paper, we have explored the dynamics of an optomechanical system with ultracold atoms between the optical cavities. Within the framework of non equilibrium Dicke model, we presented the rich phase portrait of attractors, including regimes of coexistence and persistent oscillations. We conclude from the analytical methods that the optomechanical system remains handy over an optical system in terms of control over phase transition and dynamical phase regions. The cantilever was found to be enhancing the coexisting region to a large extend and the persistent oscillation regime predicted the existence of limit cycle that prohibits reaching any stable state even in very long duration experiments. To study the system further, we added an external mechanical pump and found the external pump enhancing both the SRA and SRB phases thereby predicting an enhancement even in coexisting regions. We thereby predict a system that alters the phase transition in a Dicke model through a simple and effective process. Such system can also be used to study the dynamical entanglement in different regimes in the presence and absence of mechanical pump which can be used as a tool to selectively modify and alter the entanglement \citep{45} between the modes.

\section{Acknowledgements}
	
Aranya B Bhattacherjee acknowledge financial support from the Department of Science and Technology, New Delhi for financial assistance vide grant SR/S2/LOP-0034/2010.

\end{document}